# Experimental Demonstration of an On-Chip CMOS-Integrated 3T-1MTJ Probabilistic Bit - A P-Bit

Xuejian Zhang, John Arnesh Divakaruni Daniel, Neil Dilley, Zhihong Chen, Joerg Appenzeller

Birck Nanotechnology Center

Elmore Family School of Electrical and Computer Engineering

Purdue University

West Lafayette, Indiana, USA

zhan2808@purdue.edu

## Abstract

Ongoing semiconductor scaling challenges and the rise of neuromorphic computing have sparked interest in exploring novel computing schemes to achieve higher power efficiency and computational capabilities. Probabilistic computing is one candidate that promises low power consumption, a superior ability to solve probability-encoded computational problems, and the ease of integration with existing CMOS technology. A basic building block of this scheme is the probabilistic bit (P-Bit), which utilizes a novel device such as a stochastic magnetic tunnel junction (sMTJ) to generate tunable randomness by nature. This work presents the first experimental demonstration of a fully CMOS-integrated sMTJ-based P-Bit, capable of generating rail-to-rail stochastic output with a mere collection of 3 transistors + 1 sMTJ. Furthermore, simulations also confirm this P-Bit's functionality in configurable AND/OR probabilistic logic circuit. The demonstration of such a P-Bit paves the way towards realizing monolithic large-scale probabilistic computing architectures on CMOS chips.

## Introduction

Given the current stagnation in scaling down traditional silicon-based complimentary metal-oxide semiconductor (CMOS) devices, many turn to novel computational schemes to further advance computing power. Probabilistic computing, which greatly overlaps with stochastic computing in terms of hardware realization, is one such scheme that has shown its promise in enabling efficient computation of complicated problems at a low-energy cost by carefully programming them so to be solved in a probabilistic manner [1]-[17].Central to probabilistic computing scheme is the probabilistic bit (P-bit), a compact building block composed of just three transistors and a stochastic magnetic tunnel junction (sMTJ), enabling construction of more-complex architecture for various applications [14], [18]-[25]. Harnessing the randomness from a novel device such as a stochastic MTJ (sMTJ), each P-bit generates a stochastic output voltage fluctuating between high and low voltage levels, with the probability of being in either state controlled by an input voltage.

An MTJ is a two-terminal device that conducts currents depending on the magnetization angle between the two ferromagnets that are key parts of the MTJ. When the magnetization of the two ferromagnets is anti-parallel, the spin-filtering oxide layer between the ferromagnets passes less current (high resistance state), while a larger current is passed through the MTJ when both ferromagnets' magnetization is parallel (low resistance state) [26], [27]. Transitioning between these two states, especially when they are separated by a low energy barrier, can be visualized in terms of an energy landscape as shown in Fig. 1(a). The relative difference between high and low resistances is known as the tunnel magnetoresistance (TMR), defined as:

$$TMR = (R_{high} - R_{low})/R_{low} \quad (1)$$

Due to spin-transfer torque (STT) effect, passing currents with different polarities through the MTJ can change the magnetization of the "free" ferromagnetic layer (top CoFeB layer in our stack) relative to the fixed layer (bottom CoFeB layer), in this way changing the resistance of the MTJ [28], [29]. It is this tunability in combination with the permanent magnetic moments of the ferromagnet layers at room temperature that enables low-energy magneto-resistance random access memory (MRAM) technology [30]. Beyond memory applications, it has been demonstrated that stable MTJs can be used to create streams of random numbers by intentionally making them unstable using external stimulus [16], [31]-[35]. However, the external circuitry to enable the desired stimulus comes at the expense of additional power consumption, and the quality of this type of random number generator is limited.

Inherently stochastic MTJs (sMTJs), which do not operate under current high enough to induce STT effect, enable asynchronous computation at reduced power consumption, compared against those requiring external stimulus which is usually clocked and synchronous [12], [14], [36]-[40], and it is the key of demonstrated P-Bits in this work. As shown in Fig. 1(a) and (b), a properly designed sMTJ allows for random fluctuations between its two resistance states to occur at room temperature with 50-50 chance of staying in either state. Depending on the dimension and material composition of the sMTJ, the average time to toggle between resistance states (dwell time) can be tuned down to nanoseconds [41], [42], [43].

Shown in Fig. 1(c), the P-Bit circuit design consists of an sMTJ together with three transistors, two of which constitute an inverter [18]-[20], [44]. Ideally, the P-Bit's outputs at each input voltage average to a sigmoidal curve as shown by the orange line in Fig. 1(d). It is important that the sMTJ fluctuates with a 50-50 chance of landing in either high or low resistance state, as the P-Bit circuitry relies on this 50-50 randomness to generate a symmetric sigmoidal output curve [19], [45]. To ensure the sigmoidal curve is centered at $V_{DD}/2$, the NMOS in series with the sMTJ has to be designed to match the resistance of sMTJ. Although beyond the scope of this work, the inherent random fluctuations of such P-Bit also make it an excellent candidate for true random number generation (TRNG) [36], [40], [42], [46], [47], which has been finding critical applications in fields such as cryptography, secure communications, hardware security, and emerging machine learning paradigms that require high-quality randomness.

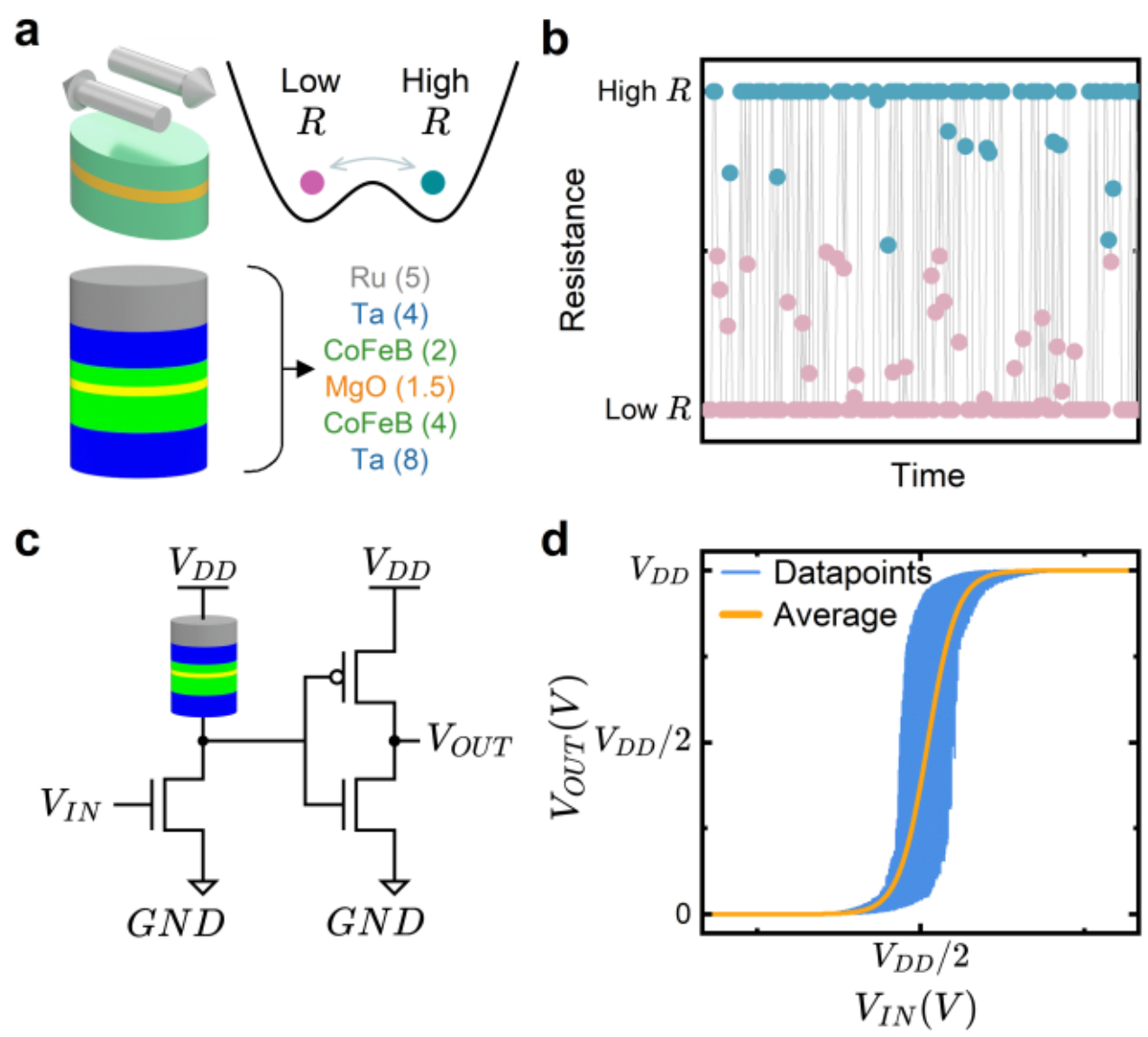

Fig. 1: Concept of an sMTJ and P-Bit based on it. (a) In-plane-anisotropy stochastic MTJ's operation in atmosphere can be analogized by the state jumping between two lowest energy wells representing the two resistance states. The MTJ stack in this work is deposited in the following composition: Ru(5)/Ta(4)/CoFeB(2)/MgO(1.5)/CoFeB(4)/Ta(8) from top to bottom with all numbers in units of nanometer. (b) Resistance of sMTJ ideally jumps between two resistance states in a true random fashion. This process is presumed to occur even without any current or voltage applied to sMTJ. (c) Schematic of a P-Bit, which consists of an sMTJ and three transistors, two of which form an inverter. The sMTJ is connected in series with one NMOS transistor, which converts stochastic resistance into voltage signal. This voltage signal is then amplified by a CMOS inverter consisting of a PMOS and an NMOS transistor. (d) Simulation results showing the behavior of a P-Bit. The input voltage is varied to tune the output fluctuations and the average output voltage level. A sigmoidal curve is observed in the average output vs input voltage plot.

This work presents for the first time CMOS-integrated P-Bits based on one sMTJ and three CMOS field effect transistors (FETs) on chip, as a milestone in probabilistic computing. Building on our successful demonstration of a P-Bit core consisting of an sMTJ and a transition metal dichalcogenide (TMD) transistor [45], this work showcases a complete P-Bit on CMOS employing back-end-of-line-(BEOL)-compatible fabrication processes. In addition, simulations of a probabilistic computing circuitry, i.e. an invertible AND/OR gate, using experimental P-Bit characteristics confirm the functionality of the demonstrated P-Bits in this work when placed in a probabilistic computing circuit. Such demonstration is the first of its kind in a fully integrated CMOS + sMTJ setting, which opens up possibilities for future more complex probabilistic computing architectures to be effectively integrated with existing CMOS technology.

## Discussion

### A. Characterization of a Stochastic MTJ (sMTJ)

MTJs are fabricated on top of a CMOS tape-out using a series of lithography, etching, and deposition processes with the fabrication details described in the Methods section. An SEM image of an example sMTJ pillar is shown in Fig. 2(a). The elliptical pillar has a major axis diameter of 130nm and a minor axis of 80nm, with its long axis aligned to the in-plane easy axis of the free layer. An external magnetic field along the long axis is still necessary in this pillar design since the dipolar field from the MTJ reference (fixed) layer tends to bias the free layer to one direction, and the applied external magnetic field cancels out this bias field so that the free layer can switch stochastically.

Fig. 2(b) shows an example of an sMTJ pillar resistance as a function of external magnetic field. Between around -7.5 and -6.9mT (called in the following the "stochastic window") resistance fluctuations are clearly visible. The probability of this device operating in the anti-parallel (AP) high-resistance state decreases with increasingly stronger B-field. From the measurement we also extract a magnetic field of -7.22mT at which the likelihood of finding the sMTJ in its AP or P state is equal, i.e. the so-called 50-50 point.

A representative oscilloscope measurement at a fixed in-plane magnetic field of 3.2mT is shown in Fig. 2(c), where the resistance of another sMTJ is observed to stochastically switch equally between high and low resistance states colored in teal and pink respectively. The TMR of this sMTJ is measured to be 14.5% and its base resistance in the parallel (low resistance) state is 27.6 kΩ. Fitting these resistance measurement data with an auto correlation function (ACF) results in a dwell time of 4.2 ms, which is the average time the sMTJ remains in one state before switching into the other state. Since the three CMOS transistors in our P-Bit can operate at frequencies beyond GHz levels, the dwell time of the sMTJ entirely determines the speed of the P-Bit, defining the shortest sampling interval to guarantee the P-Bit's output randomness.

While the device in Fig. 2(c) is not the fastest sMTJ reported in this work, it is discussed here first because the associated P-Bit exhibits the most ideal sigmoidal output characteristics. Note that faster switching can be enabled by changing the MTJ design, since the energy barrier between the two MTJ states is determined by the volume and anisotropy of the free layer [12], [48], [49]. Moreover, recent work has experimentally shown nanosecond-level switching in sMTJs [41]-[43], [50], which implies that higher-throughput P-Bits will be available in future implementations.

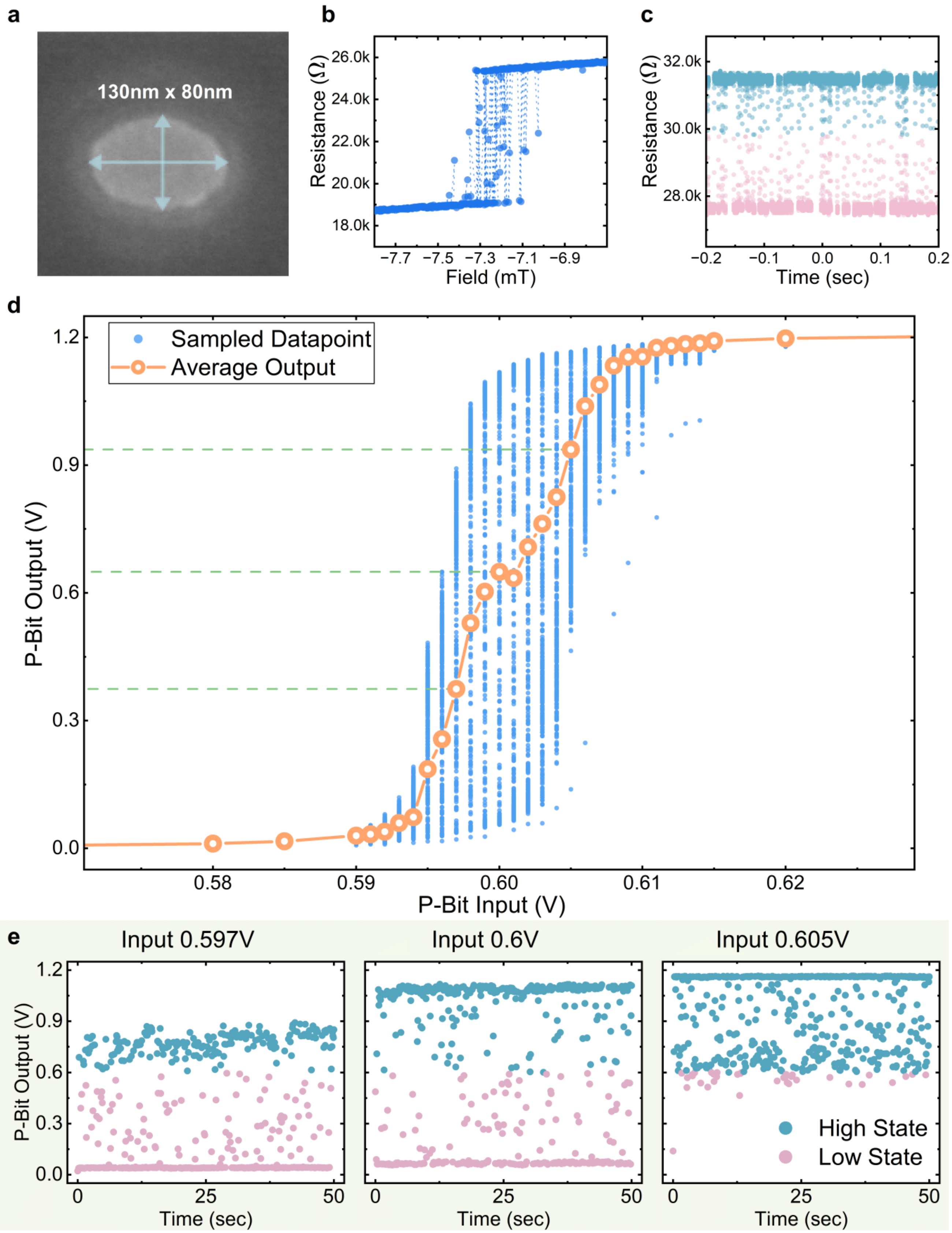

Fig. 2: Example P-Bit measurement results. (a) SEM image of a fabricated sMTJ pillar. During operation, a biasing magnetic field is usually applied along the long axis, i.e. the in-plane easy axis. (b) Resistance of sMTJ pillar vs. minor magnetic field loop for an sMTJ pillar. The stochastic resistance can be observed over a small range of biasing B field. (c) Resistance of an sMTJ vs. time measured with an oscilloscope at 50-50 point. Pink and teal dots represent datapoints treated as low/high resistance states. (d) P-Bit output (1sMTJ + 3FETs) characterized using a parameter analyzer, with 500 datapoints sampled at each input voltage for a duration of 50sec. Small blue dots represent the real-time sampled output voltage for a fixed input voltage. Hollow orange circles are average output voltages. The output is well-centered around $V$DD/2 and shows almost rail-to-rail stochastic behavior for sufficiently high and low input voltages respectively. (e) Voltage output of the same P-Bit at three different inputs, showing the tunability of the output voltage distribution.

## B. Experimental Demonstration of a CMOS-Integrated P-Bit

When placed inside the P-Bit circuit, the sMTJ shown in Fig. 2(c) acts as a source of randomness. The circuit output of the P-Bit is measured using a semiconductor parameter analyzer. More details about the measurement setup can be found in the Methods section. Fig. 2(d) shows measurement results of the world's first CMOS-integrated sMTJ-based P-Bit using the TSMC 180nm node. Small blue dots represent the sampled datapoints at fixed input voltages. Similar to measurement Fig. 2(c), a constant magnetic field is applied along the easy axis.

In Fig. 2(d), a P-Bit's output is observed to stochastically switch almost rail-to-rail between $V_{DD}$and GND when the P-Bit input is at $V_{DD}$/2. As the input voltage sweeps from 0.58V to 0.62V, the output voltage distribution gradually biases from low to high, while the average output voltage follows a sigmoidal shaped curve. Fig. 2(e) displays three collections of sampled datapoints at different input voltages to illustrate the tunability of the output voltage distribution.

When the input voltage is at 0.597V or 0.605V, the output of the P-Bit is biased low or high, respectively as desired. On the other side for $V_{IN}$=$V_{DD}$/2=0.6$V$, the output is evenly distributed between the two states. The symmetry around $V_{DD}/2$ is important to allow multiple P-Bits to be cascaded in circuits. Properties such as the tunability, rail-to-rail output capability and the symmetrically-centered sigmoidal output curve make this P-Bit a good candidate for implementing larger probabilistic circuits.

Prior simulations and experiments [45], [51] have already shown that the sigmoidal curve of a P-Bit is highly sensitive to the 3 transistors and their matching with the sMTJ. During the CMOS integration, it is critical at the design stage to set an adequate $V_{DD}$to be compatible with the sMTJ layout. For the sMTJ film developed in this work, the MgO thickness is tuned to be sufficiently thick to suppress spin-transfer torque (STT) from occurring at any voltage between 0 and $V_{DD}$, which would impact the 50-50 point by making it current-bias dependent. For our 1.5nm thick MgO layer, we found that the specified supply voltage of 1.8V for the 180nm TSMC node resulted in currents which were too large. To prevent STT from impacting our P-Bit performance all experiments were performed at a supply voltage of $V_{DD}$= 1.2V.

A faster P-Bit was successfully fabricated on another similar CMOS tape-out, though its sigmoidal output curve is not as ideal as the one shown in Fig. 2. This faster P-Bit's sMTJ is individually measured to exhibit a dwell time of 68.9$\mu s$, as shown by the oscilloscope trace in Fig. 3(a). It is worth noting that the stochastic TMR (sTMR) is measured to be 30%, higher than the 14.5% sTMR of the previous sMTJ. The output characteristics of this faster P-Bit is shown in Fig. 3(b). The output is close to rail-to-rail in the fluctuation dominated region, but the sigmoidal curve is clearly distorted. Though higher sTMR helps creating more pronounced rail-to-rail fluctuations at the P-Bit's output, it can lead in particular to "plateaus" as discussed in [45]. Furthermore, we believe that other non-idealities of the sigmoidal shaped curve are related to a narrow stochastic window of ~0.2mT (as shown by the minor magnetic field loop in Fig. 3(c)) for the specific sMTJ at the heart of this P-Bit, making it hard to maintain a consistent 50-50 resistance fluctuation distribution during P-Bit measurement.

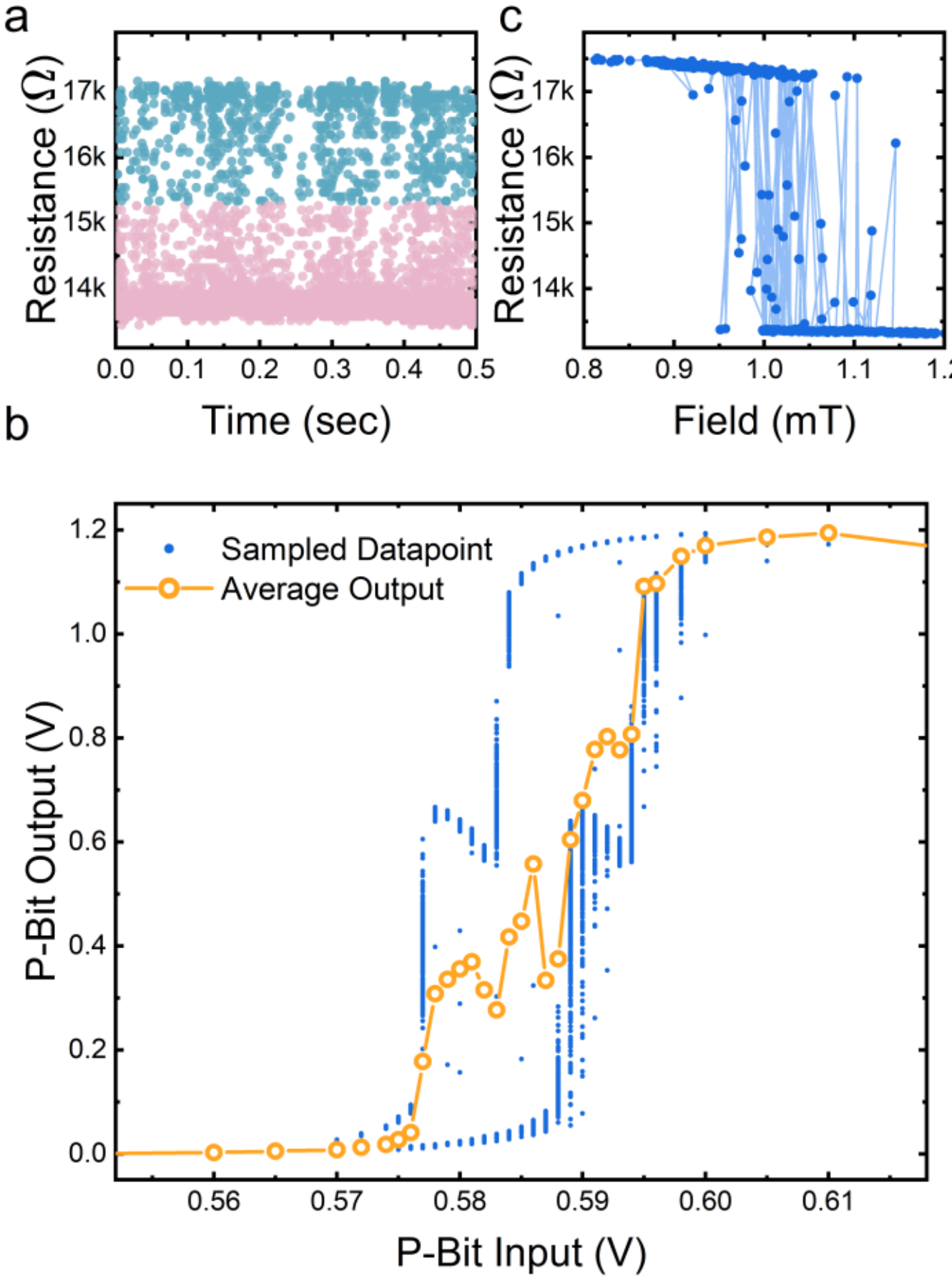

Fig. 3: Another P-Bit based on a faster-switching sMTJ. (a) Oscilloscope measurement of sMTJ's resistance over time. Pink dots denote resistance values considered low and teal as high respectively. The dwell time of this sMTJ is calculated to be 68.9$\mu s$. The resistance values are slightly shifted to remove DC offset according to the stochastic signal measured using an SMU. (b) Output characteristics of this P-Bit, measured using the same setup as in Fig. 2. The output still yields substantial output voltage fluctuations. However the sigmoidal curve reveals non-idealities that we attribute to: 1) a higher stochastic TMR, which we discussed in [45], can result in "plateaus" in the sigmoidal output/input characteristics of the P-Bit and 2) a very narrow stochastic window for the associated sMTJ making it harder to maintain 50-50 fluctuation, which causes the averaged P-Bit output values to not follow a consistently increasing trend as the P-Bit input increases. (c) Minor magnetic loop on the sMTJ. The stochastic window is as narrow as ~0.2mT.

## C. Performance Metrics Comparison and Projection

The average power consumption of the two demonstrated P-Bits (P3) is plotted against the sampling throughput determined by the sMTJs' dwell time, as shown in Fig. 4. The comparison also includes two other works (P1 and P2) that have successfully demonstrated probabilistic bits based on PCB-level integration [52], [53]. By realizing the low power consumption benefits from a monolithic on-chip integration with CMOS, the two CMOS P-Bits from this work show significant improvements in power efficiency compared against predecessors. The average power drawn by our CMOS P-Bits is measured to be around 52 to 55$\mu W$, while the PCB-level P-Bits consume a total power of 8 to 10mW (this includes the whole package of sMTJ and transistors to generate rail-to-rail output).

The fastest sampling throughput of a single P-Bit in this work is $\sim 1.4 \times 10^{-5}$ flips/ns, as determined from its dwell time of 68.9$\mu s$. However, substantially higher throughput of 1 flip/ns is achievable with demonstrated nanosecond-level switching times in sMTJs [41]-[43], [50]. As the throughput of a P-Bit is determined by the switching frequency (inverse of dwell time) of the sMTJ, because of the fast switching of CMOS inverters, one can predict (P4).

From the CMOS P-Bits demonstrated here, our power consumption projection (P4) makes the following assumptions: (1) the average resistance of an sMTJ is around 100kΩ, (2) the CMOS process node supports a core voltage of 0.9V, and (3) the load of an inverter using said node is 1fF. With these numbers, we project that an sMTJ-based P-Bit can generate a throughput of 1 flip/ns at a power consumption of $4.9\mu W$ of which $\sim$800nW is attributed to the inverter, plotted in Fig. 4.

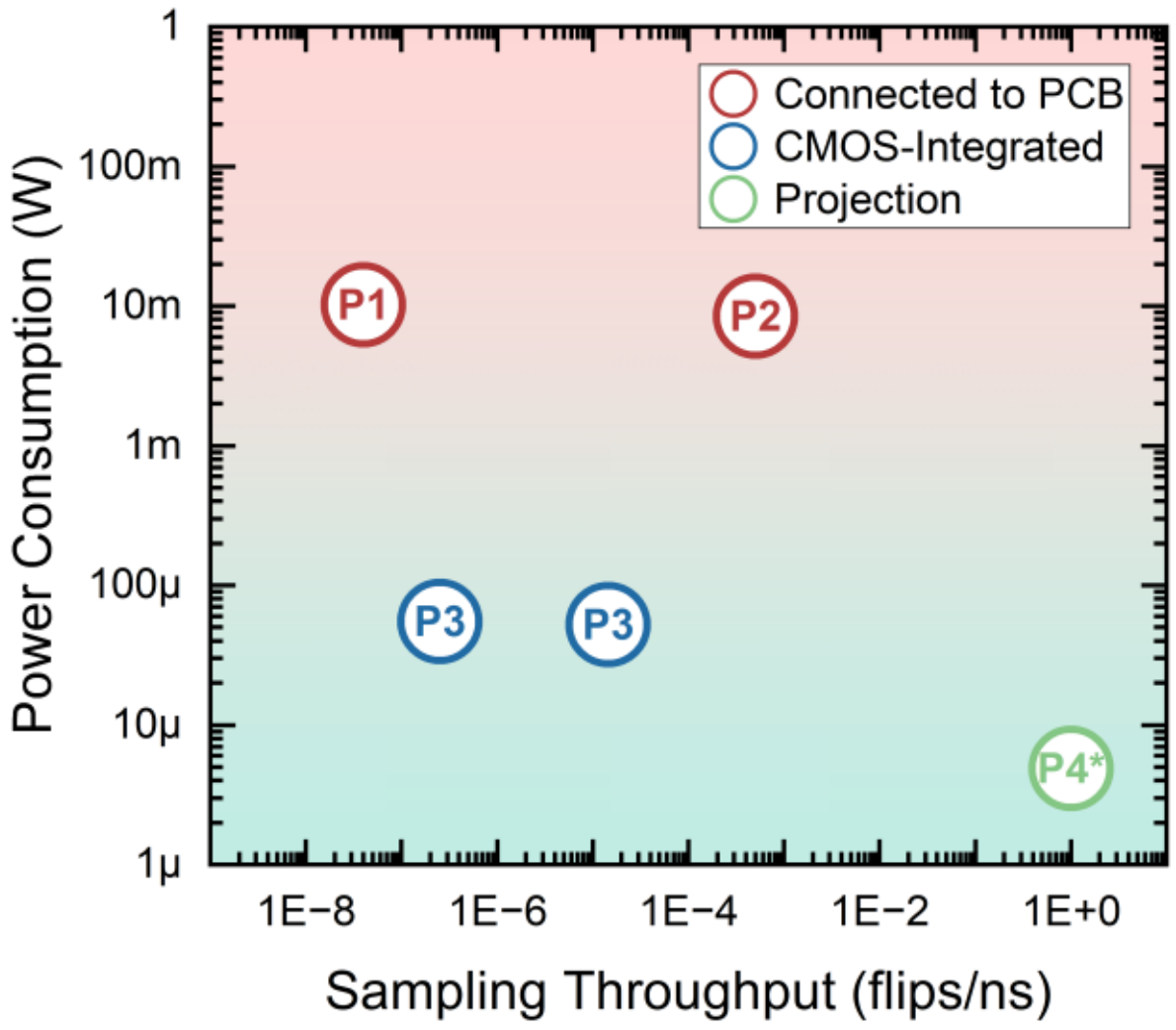

Fig. 4: Performance metric of CMOS-integrated P-Bits. Power consumption of each demonstrated P-Bit is measured at an input voltage level where the output of the P-Bit fluctuates equally between $V_{DD}$ and GND. At this voltage level, the average power consumption is calculated by measuring the average current drawn from $V_{DD}$. The switching frequency is first determined by the inverse of the dwell time of the associated sMTJ, and the throughput is then calculated as the number of flips generated per nanosecond. Performance metrics collected from our two CMOS P-Bits are compared against previous work on probabilistic circuits (P1 refers to [52] and P2 to [53] respectively) whose P-Bits are of similar structure. The throughput calculation assumes that the CMOS FETs are not the limiting factor. Based on experimentally observed switching speeds of 1ns and considering lower power consumption in advanced CMOS nodes if compared to the one used here, P4 is our projected metric leading to a throughput of 1 flip/ns at a power consumption of $4.9\mu W$. This projection is half of the one shown in [53].

## D. Simulation of Probabilistic Logic Circuit with Demonstrated P-Bit

Next, simulations have been performed to verify the functionality of the demonstrated P-Bits at the circuit level. The simulated circuit is based on the tunable AND/OR probabilistic logic gate proposed in [19], [44], which consists of three P-Bits and a network of resistors as weights or synapses, shown in Fig. 5(a). The network of weights forces the P-Bits to output combinations corresponding to AND/OR logic gates, by shaping an energy landscape to favor correct combinations at lowest energy levels. This probabilistic logic gate can compute in both forward and reverse directions - meaning that it can identify allowed input combinations when the output is clamped to a certain state. The experimental P-Bit demonstrated in Fig. 2(d) is modeled in Virtuoso Cadence by feeding in randomly generated sMTJ resistances based on real measurements. The randomly generated sMTJ resistances follow the same statistical distribution as the measured resistances. The transistor size for simulation is adjusted so that the simulated P-Bit follows the demonstrated P-Bit's characteristics in Fig. 2(d). The resulting average output sigmoidal curves in Fig. 5(b) show good agreement between simulation and experiment.

For the chosen circuit level simulations, the same circuit's weights network is biased in two different ways to perform the OR gate and AND gate operations separately [19]. During the OR gate simulation, the output node C is clamped to a low state 0 for one run of simulation and the voltage readings at nodes A and B are recorded and digitized to create the histogram shown in Fig. 5(c). The peak of the histogram at ABC = 000 in OR mode operation indicates that both A and B can only output 0 when C is clamped to 0. Meanwhile, if node C is clamped to a high state 1 for another run of simulations, the histogram of digitized ABC output will show almost equal probability for ABC = 001, 010 and 111. The result described above matches with the truth table of a logical OR gate. The same procedure is applied to simulate an inverted AND gate operation, which shows almost equal probability of ABC = 000, 001 and 010 when node C is clamped to 0, and a single peak at ABC = 111 when node C is clamped to 1. This result is also consistent with the truth table of a logical AND gate.

Simulation result in Fig. 5(c) shows imperfections in two cases: OR mode inverted operation where C is clamped to 1 and AND mode inverted operation where C is clamped to 0. In both cases, the histograms are supposed to show three peaks of equal probability, while the simulation results show slightly different peak heights. These imperfections are likely attributed to the non-ideal sigmoidal curve of the simulated P-Bit, which suffers from the “plateau” effect as discussed in [45]. Despite these imperfections, the overall result confirms that the demonstrated P-Bit is a functional building block for probabilistic logic circuits, bringing realization of probabilistic computing on CMOS chips closer to reality.

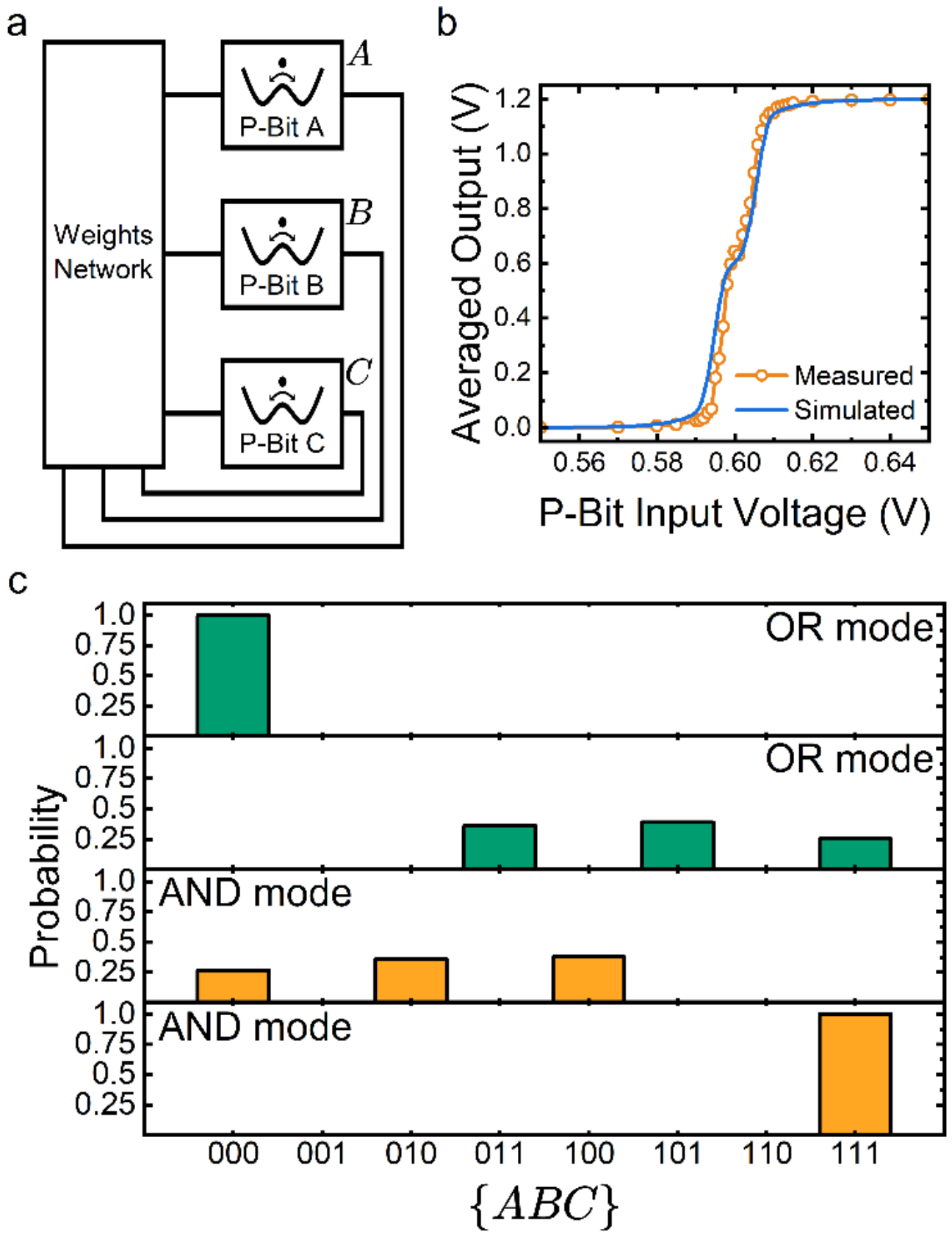

Fig. 5: Probabilistic AND/OR gate simulation performed with demonstrated P-Bit's characteristics. (a) The circuit of a configurable AND/OR gate consists of three independent P-Bits, together with a weight network [19] (b) Comparison between simulated P-Bit output characteristics and experimentally demonstrated P-Bit's output characteristics. The sigmoidal output of simulated P-Bit is fitted to match that from experiment. (c) Simulation result of the inverted AND/OR gate with the output node C clamped to either 0 or 1 during each operation mode. The histogram of digitized ABC output combinations shows that the inverted operation of AND/OR gate follows the truth table of AND/OR logic gate.

## Conclusion

This paper demonstrates two working P-Bits successfully integrated for the first time with CMOS tapeouts. The experimentally-demonstrated P-bits are capable of generating almost rail-to-rail stochastic output voltages; the particular one illustrated in Fig. 2 shows a total power consumption of 55.2uW at 1.2V supply voltage, while its output is tunable and perfectly centered at $V_{DD}/2$. Performance comparison between CMOS-integrated P-Bits and previously published results highlight the potential for sMTJ-based P-Bits to achieve both power-efficiency and novel computational capabilities. Circuit-level simulations using experimental P-Bit data confirm the functionality of the demonstrated P-Bit in a probabilistic AND/OR logic circuit. Future work will focus on optimizing the fabrication process to improve the performance metrics of the P-Bits, as well as exploring the potential of integrating multiple P-bits to create larger-scale probabilistic circuits towards realizing probabilistic computing in a monolithic chip.

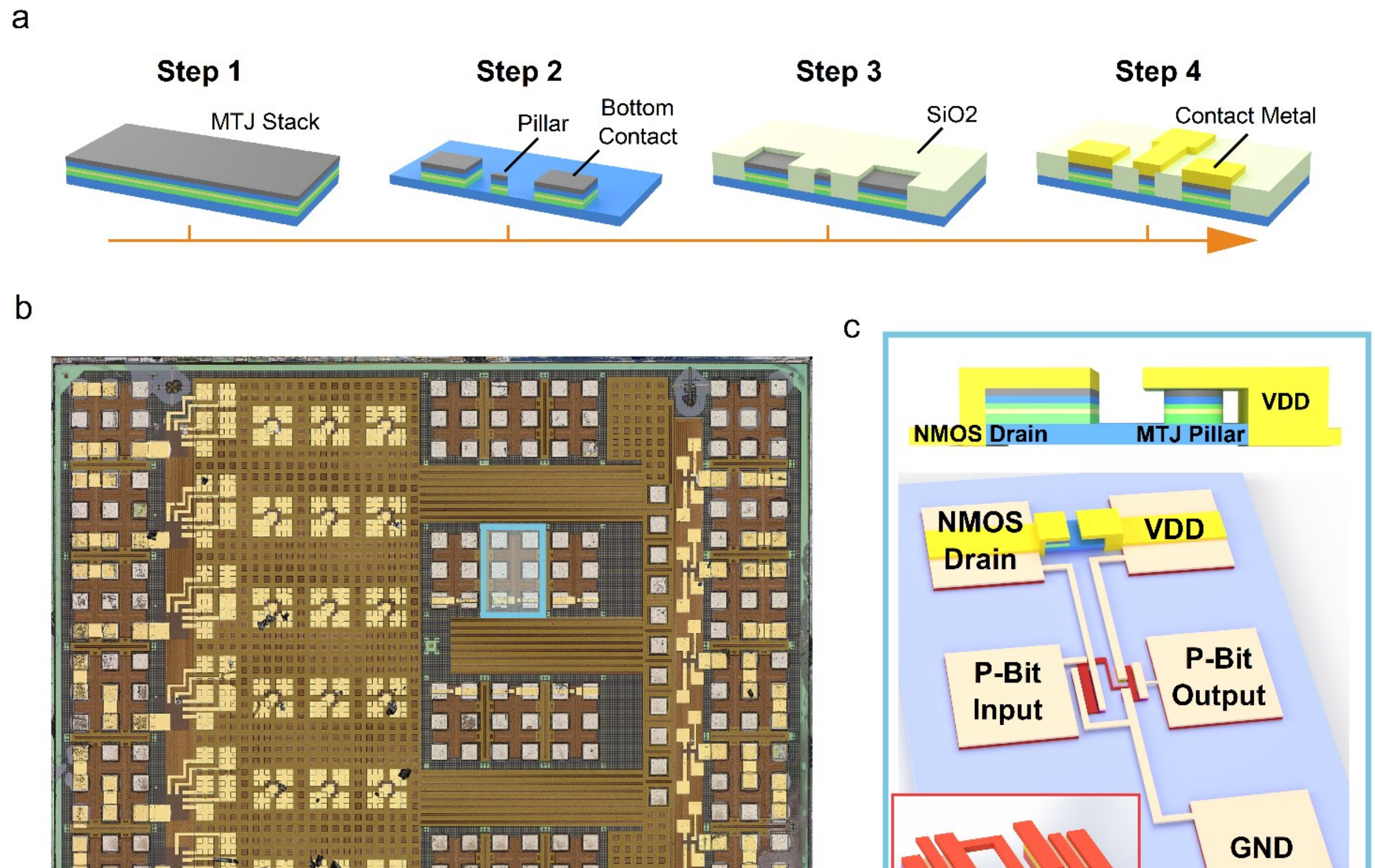

Fig. 6: Fabrication process of sMTJs on TSMC 180nm node tape-out. (a) The cross-sectional view shows the entire process flow. The multi-layer MTJ stack is sputter-deposited on the substrate, followed by etching to first create a larger MTJ block and then an elliptical pillar in the center of this block. Immediately after the etching process, a layer of SiO2 is deposited to ensure that the top contact to the pillar is not connected to any other terminal. Finally, Ti/Au is deposited on top to create electrical connection to both the top Ta/Ru and bottom Ta layer of the MTJ pillar. (b) The tape-out chip on which P-Bits are fabricated, with one successfully integrated P-Bit indicated by the blue square. (c) 3D illustration of P-Bit circuitry, with the sMTJ pillar fabricated between $V_{DD}$ and the NMOS FET drain node. The inset showcases the placement of CMOS transistors within the P-Bit circuitry, where the red blocks represent poly-silicon or metal layers for FET gates and interconnects.

## Methods

This work contains three major parts - Deposition of MTJ film, MTJ fabrication, CMOS design and integration. The entire process is performed by our research group and each part has been carefully optimized before fabrication of the P-Bit on CMOS.

### A.MTJ Film Deposition

MTJs demonstrated in this work are built from sputtered films that consist of (from bottom to top) Ta(10nm)/CoFeB(4nm)/MgO(1.5nm)/CoFeB(2nm)/Ta(4nm)/Ru(5nm). MgO was deposited using RF magnetron sputtering guns and the remainder of the stack was deposited using DC magnetron sputtering in the same chamber. The rates of material deposition are calibrated beforehand employing atomic force microscope (AFM) measurements.

The bottom Ta layer functions as a bottom seeding layer and helps reduce the dead layer thickness in the CoFeB layer grown atop. Both CoFeB ferromagnetic layers act together to control the TMR. The bottom

CoFeB layer is used as a reference (fixed) layer due to its higher coercivity, while the top CoFeB layer behaves as a free layer. The MgO layer acts as a spin-filtering tunneling layer and is key for controlling the base resistance and TMR. The thickness of the MgO layer is optimized to 1.5nm for resistance-matching purposes, while this thickness also avoids dielectric breakdown during operation. The top Ta/Ru layer is used to prevent oxidation of the CoFeB layer underneath, where the Ru layer on top of Ta prevents oxidation of Ta layer itself.

B.MTJ Fabrication Process

Since the sMTJs are fabricated on top of a CMOS tape-out, the entire fabrication process must be back-end-of-line (BEOL) compatible to avoid negatively impacting the function of the designed circuits underneath. Test fabrication of sMTJs is first performed on Si/SiO2 substrates with no underlying CMOS circuits and then the working process flow is transferred onto the CMOS tape-out.

As illustrated by Fig. 6(a), sputter deposition is first applied to grow the desired sMTJ films on the CMOS substrate. Next, E-Beam Lithography (EBL) and Ar-ion milling are utilized in step 2 to define and etch the sMTJ pillars (shaped below 200nm in diameters) and comparatively larger (5x10um) sMTJ stacks as contacts to the bottom of MTJ pillars. The 5x10um sMTJ stacks do not contribute to magnetic TMR because pinholes in MgO layer of this size result in low resistances. Immediately following the etching step, SiO2 is deposited in step 3 to electrically isolate the MTJ pillar and bottom contact stack. Step 4 employs evaporation-deposition of 40nm Ti and 120nm Au to form electrical connection with both the top and bottom electrodes of the sMTJ pillars.

C.CMOS Design, Integration and Measurement

The CMOS tape-out is designed using the commercial mixed-signal 180nm process node from TSMC. The P-Bit circuit is designed using Virtuoso Cadence and the layout is verified to pass design rule and layout vs. schematic checks. Each tape-out chip contains internally connected transistors allowing a maximum of 24 P-Bits to be demonstrated. Large Al top metal pads are placed in grids of 2x3 or 3x3 blocks, and the vacant areas between these blocks are intentionally designed to suit the sMTJ fabrication. After the tape-out is received, the MTJ fabrication process as described in the previous section is applied to create MTJs on top of CMOS circuits to form P-Bits, shown in Fig. 6(b). A total of 110 sMTJs are fabricated on the tape-out, with each elliptical-shaped MTJ sized for a balance between yield and resistance.

During measurement, the CMOS chip with P-Bits is placed in a Lakeshore CRX-EM-HF probe station, which allows for applying an external magnetic field along the in-plane easy axis of the sMTJ. Oscilloscope measurements are directly performed on sMTJs with Keithley 6221 current source and Keysight MSOX-3014T oscilloscope (with probes landed on NMOS Drain and VDD pads as shown in Fig. 6(c)). The current source outputs a DC current of 10uA and the oscilloscope is used to measure the voltage from sMTJs at a sampling frequency of 100kHz. The measurement setup of a P-Bit is shown in Fig. 6(c), where an Agilent 4156C parameter analyzer is connected to the P-Bit circuit through four probe tips ($V$DD, P-Bit Input, P-Bit Output and GND). The NMOS FET drain node is not directly measured as it is only used to connect to the bottom node of the sMTJ and the inverter. Detailed placement of transistors and interconnects is also shown in Fig. 6(c), which are directly connected to Al pads for probing.

## Acknowledgements

This work is supported by the National Science Foundation under grant number DMREF-2324203. The authors would like to thank Purdue University Birck Nanotechnology Center for providing access to the fabrication facilities and MUSE Semiconductor for providing access to CMOS tape-out service through TSMC.